\documentstyle[12pt,fleqn,psfig]{article}
\newcommand{\nwc}{\newcommand}

\newcommand{\be}{\begin{equation}}
\newcommand{\ee}{\end{equation}}
\newcommand{\nn}{\nonumber}
\newcommand{\beba}{\begin{equation}\begin{array}{lcl}}
\newcommand{\eaee}{\end{array}\end{equation}}
\newcommand{\bea}{\begin{eqnarray}}
\newcommand{\eea}{\end{eqnarray}}
\newcommand{\ba}{\begin{array}}
\newcommand{\ea}{\end{array}}

\newcommand{\ns}{\normalsize}
\newcommand{\refs}[1]{(\ref{#1})}
\newcommand{\ler}{\stackrel{\scriptstyle <}{\scriptstyle\sim}}
\renewcommand{\Re}{{\rm Re}}
\renewcommand{\Im}{{\rm Im}}
\nwc{\ra}{\rightarrow}
\nwc{\lra}{\longrightarrow}
\nwc{\lera}{\leftrightarrow}
\nwc{\lolera}{\longleftrightarrow}
\nwc{\pa}{\partial}
\nwc{\pri} {^{\prime}}
\nwc{\dpr} {^{\prime\prime}}
\def\a{\alpha}
\def\b{\beta}

\def\c{\chi}
\def\d{\delta}
\def\e{\epsilon}
\def\ve{\varepsilon}
\def\f{\phi}
\def\vf{\varphi}

\def\m{\mu}
\def\n{\nu}

\def\p{\pi}

\def\r{\rho}
\def\s{\sigma}

\def\D{\Delta}

\def\ch{{\cal H}}
\def\cl{{\cal L}}
\def\cc{{\cal C}}
\def\bx{{\bf x}}
\def\bk{{\bf k}}
\def\bn{{\bf n}}
\def\vfb{{\mbox{\boldmath $\vf$}}}
\def\cb{{\mbox{\boldmath $\c$}}}
\def\zz{\relax{\sf Z\kern-.3em Z}}
\def\ZZ{\relax{\sf Z\kern-.4em Z}}
\def\ZZZ{Z\kern -0.28em Z}
\def\CC{{\rm \kern .25em
             \vrule height1.4ex depth-.12ex width.06em\kern-.31em C}}
\begin{document}
\begin{titlepage}
\title{{\large\bf Decoherence in Pre--Big--Bang Cosmology}\\
                          \vspace{-4cm}
                          \hfill{\ns TUM-HEP 240/96\\}
                          \hfill{\ns SFB-375/87\\}
                          \hfill{\ns hep-th/9603167\\[.1cm]}
                          \hfill{\ns March 1996}
                        \vspace{2cm} }

\author{Andr\'e Lukas\setcounter{footnote}{0}
        \thanks{Email : alukas@physik.tu-muenchen.de}
        \setcounter{footnote}{3}
        \thanks{Address after March 1996~: Department of
        Physics, University of Pennsylvania,
        Philadelphia, PA 19104, USA.}\\[2mm]
        \setcounter{footnote}{6}
        {\ns and}\\[2mm]
        Rudolf Poppe\thanks{Email : rpoppe@physik.tu-muenchen.de}\\[1cm]
        {\ns Physik Department}\\
        {\ns Technische Universit\"at M\"unchen}\\
        {\ns D-85747 Garching, Germany}\\}

\date{}
\maketitle

\begin{abstract} \baselineskip=6mm
We analyze the quantum cosmology of the simplest pre--big--bang model 
without dilaton potential. In addition to the minisuperspace variables
we include inhomogeneous dilaton fluctuations and determine their wave
function on a semiclassical background. This wave function is used to
calculate the reduced density matrix and to find criteria for the loss 
of decoherence. It is shown that coherence between different backgrounds 
can always be achieved by a specific choice of vacua though generically
decoherence can be expected. In particular, we discuss the implications 
of these results on the ``exit problem'' of pre--big--bang cosmology.
\end{abstract}

\thispagestyle{empty}
\end{titlepage}


\section{Introduction}

Most approaches to string cosmology start from an effective theory
of the dilaton and the moduli which move in an ordinary
Robertson--Walker background defined in the {\em Einstein frame}.
Unfortunately, this ``conventional'' approach has a number of
serious drawbacks. It appears
to  be difficult to stabilize the dilaton in its minimum and to generate
an inflating phase~\cite{br_stein}. Moreover, the energy density stored
in the oscillating moduli typically dominates the energy density of 
the universe~\cite{mod_prob} thereby destroying the standard big bang
nucleosynthesis scenario.\\

These problems have motivated the search for alternative frameworks of string
cosmology and the development of ``pre--big--bang'' cosmologies~\cite{pbb}
in particular. The essential difference of pre--big--bang models as
compared to conventional scenarios is that the space--time
metric is defined in the {\em string frame} which is conformally related
to the Einstein frame. In the former stringy symmetries
though of course also present in the Einstein frame formulation show up
in a much more direct way. In the simplest model of cosmological relevance,
consisting of a Robertson--Walker metric and a dilaton without potential,
two types of solutions arise which are related to each other by
a duality transformation acting on the scale factor $a$ like
$a\rightarrow 1/a$. This so--called scale factor duality~\cite{sfd} maps
solutions with an ordinary radiation--like behaviour of the scale factor
and increasing horizon size into solutions with superinflationary
behaviour, i.~e.~with a shrinking horizon. This phase of superinflation
is driven by the kinetic energy of the dilaton. Its unusual properties
which are so troublesome in the conventional framework
are turned into an advantage here~: They at least guarantee a phase of
inflationary expansion, the pre--big--bang phase.

{}Finally of course, after a sufficiently long period of superinflation, a
transition to a radiation dominated universe has to be generated.
In the effective theory, however, this phase is separated from the
pre--big--bang phase by a curvature singularity. This singularity
should be an artifact of the low energy effective theory and one might hope
to find a true conformal background which smoothly interpolates between the
two branches. Unfortunately, such a background is not known so far. An
alternative approach to achieve an exit from the pre--big--bang expansion is
to consider mechanisms which can be studied in terms of the low energy
effective action and come into operation before the system reaches the
region of large curvature close to the singularity. The case of a potential
as a source of branch change has been studied for the first time in
ref.~\cite{bru_ven}. It has been claimed and later proven
rigorously~\cite{olive} that such a potential cannot solve the exit problem.\\

In view of this failure in using the simplest ``classical'' extension of the
low energy theory one might try to study quantum effects
which are accessible in the effective theory. Applying the formalism of
quantum cosmology~\cite{hall} is an interesting option in this direction
which has been proposed in ref.~\cite{bento_bert}. The quantum cosmology
of a more general class of string vacua with $O(d,d)$--symmetry~\cite{odd}
which include the pre--big--bang models has been considered in
ref.~\cite{keh_luk}. For these models which, in the language of quantum
cosmology, constitute minisuperspace models (i.~e.~they consist of a finite
number of degrees of freedom), a complete set of solutions for the
Wheeler--de Witt equation can be found due to their large symmetry. A possible
application of the quantum cosmological approach to the exit problem
was first discussed in ref.~\cite{ven} where a 4--dimensional
Robertson--Walker metric and a somewhat unconventional potential depending
on the {\em invariant} dilaton has been used.\\

In this paper we are concerned with an even simpler model, namely the
aforementioned Robertson--Walker type model in 4 dimensions with a
{\em free} dilaton. Our essential point will be not just to discuss the
minisuperspace theory of that model but to go beyond mini\-superspace and
incorporate inhomogeneous fluctuations of the dilaton field. Our results
can be useful to discuss density fluctuations in pre--big--bang cosmologies
but in this work we will focus on the aspect of
decoherence~\cite{dec_gen,dec_tech,dec_hall}.\\

Along with a semiclassical behaviour of the wave function decoherence
of different classical trajectories predicted by such a wave function is
an essential condition for a classical universe. It
has been shown for several models that inhomogeneous fluctuations in the
universe can cause decoherence~\cite{dec_tech,dec_hall}. Technically
speaking, unobserved fluctuation modes e.~g.~those with wavelengths larger
than the horizon are integrated out to obtain a reduced density matrix.
This procedure might suppress the off--diagonal terms in the density
matrix thereby leading to decoherence.

The main purpose of this paper is to study the issue of decoherence for
the simplest pre--big--bang model. We consider this to be of particular
relevance for several reasons. First of all, as any
other cosmological model, pre--big--bang models treated in a quantum
cosmological way should -- at least asymptotically -- show the emergence
of classical space--time. Furthermore, it is particularly interesting
to study the mutual decoherence properties of the two branches.
Decoherence of the pre--big--bang and the ordinary branch might destroy
a quantum cosmological transition in minisuperspace. On the other hand,
loss of decoherence in some region of minisuperspace is an interesting
option to generate a branch change.

To study these questions we will start with a short review of pre--big--bang
models in section 2 which will also fix our notation and conventions.
The minisuperspace theory of the simplest model will be discussed in section 3.
We will determine the general solution of the Wheeler--de Witt equation
and identify two semiclassical wave functions which are suitable to
describe the two classical branches of the theory. In section 4 we determine
the wave function of the dilaton fluctuations on this semiclassical
background. Decoherence is analyzed in section 5 and the results and their
possible implications for a branch change are discussed in section 6.


\section{The general framework}
Starting point of our discussion is the bosonic part of the one--loop
string effective action~\cite{eff_action}
\be
 S =  \int d^4 x\; \cl = -\frac{1}{16 \pi G}\,\int\; d^4 x\;\sqrt{-g}\, 
      e^{-\vf}\left[
      R+g^{\m\n}\,\partial_\m\vf\,\partial_\n\vf
      +V(\vf)\right] \ , 
\label{action}
\ee
where $\vf$ is the dilaton and  $g_{\m\n}$ the $\s$--model metric. We have also
included a dilaton potential $V(\vf )$ originating from nonperturbative
effects. Whenever easily possible we will keep this potential throughout
the discussion. For the explicit solutions we will, however, concentrate
on the case $V=0$. Since, in this paper, we will not consider a nonvanishing
antisymmetric tensor we have dropped the corresponding term in
eq.~\refs{action}. 

Having a cosmological application in mind we stick to 4 dimensions and
use a background metric of Robertson--Walker type defined by 
\be
g_{\m\n}dx^{\m} dx^{\n} = N^2 (t)\, dt^2 - a^2 (t)\, d\bx \cdot d\bx \ .
\label{metric}
\ee
Here $N(t)$ denotes the lapse function and
$a(t)$ is the scale factor of the universe. The corresponding homogeneous
dilaton mode is called $\vf_0 (t)$. In addition to that small
inhomogeneous fluctuations $\d({\bf x},t)$ of the dilaton are taken into
account~:
\be
\vf ({\bf x},t) = \vf_{0} (t) + \d({\bf x},t) \ .
\label{split}
\ee
Within a box of volume $v=l^3$ the inhomogeneous part can be expanded into
its modes as follows~:
\be
\d({\bf x},t) = \sum_\bk\left[\vf_{\bk} (t)\, e^{i\bk \cdot \bx}
                + \vf_{\bk}^{\ast} (t)\, e^{-i\bk \cdot \bx}\right] \ .
\label{fourier}
\ee  
The sum is performed over all wave vectors $\bk = \frac{2\pi\bn}{l}\ne 0$ with
$\bn\in\ZZ^3$.

We are aware of the fact that the above choice of field configurations is
not the most general one for our purpose. E.~g.~we have not considered the
inhomogeneous scalar modes of the metric tensor which mix with the dilaton
modes even in the linear approximation~\cite{hh}. Adding these modes to the
theory would, however, complicate our analysis substantially. For simplicity,
we will, therefore, restrict ourselves to the expansion indicated above as
usually done in the discussion of decoherence in quantum
cosmology~\cite{dec_tech,dec_hall}. We do not expect our main conclusions to
depend on this particular choice.\\

For a constant dilaton potential $V_0 =$ const, the background action
$\cl_{0}$ to be obtained from eq.~\refs{action} turns out to be invariant
under the symmetry
\be
a \lra a^{-1}, \hspace{4mm} \vf_0 \lra \vf_0 - \ln a^6 \ ,
\label{sfd}
\ee
called scale--factor duality (SFD)~\cite{sfd}. This symmetry
can be viewed as a direct generalization of the
$R\rightarrow 1/R$ duality in toroidal string compactification. In the
context of more general models with homogeneous metric and nonvanishing
antisymmetric tensor it can be embedded in an $O(d,d)$ symmetry~\cite{odd}.

The above transformation law motivates the introduction of the new background
coordinates
\be
\f = \vf_0 - 3\a ,\hspace{4mm} \a=\ln a \ ,
\label{shift}
\ee
which under SFD transform in a particularly simple way~:
\be
\a \lra - \a , \hspace{4mm} \f \lra \f \ .
\label{sfd2}
\ee 
Especially using the invariant dilaton $\f$ simplifies the formulae
considerably. Let us, therefore collect the relevant ``classical''
expressions in the new basis. For the Lagrangian expanded up to second order
terms in the dilaton perturbations we get
\bea
\cl &=& \cl_0 +\sum_{k\ne 0} \cl_k \nn \\
\cl_{0} &=& \, K\, N\, e^{-\f}\,\left[\frac{1}{2 N^2}\,\dot{\a}^2 
        - \frac{1}{6 N^2}\,\dot{\f}^2 
        - \frac{1}{6}\, V_0 \right] \\
\cl_{k} &=& \frac{K}{3}\, N\, e^{-\f}\,\left[
        - \frac{1}{N^2}\left|\dot{\vf_{k}}\right|^2
        + \frac{1}{N^2}\,\dot{\f}\left(\vf_{k}\dot{\vf}_{k}^{\ast} 
              + \vf_{k}^{\ast}\dot{\vf}_{k}\right)
        + U_{k}\,|\vf_{k} |^{2} \right] \nn \ ,
\label{lagrangian2}
\eea
with the dilaton background potential $V_0 = V(\vf_0 )$, $K =3v/8\pi G$
and $k=|\bk |$. Furthermore, we have used the abbreviation
\be
U_{k} = V_{0}^{\prime}-V_{0}-\frac{1}{2}{V_0}^{\prime\prime}
        +\frac{k^{2}}{a^2} \ ,
\label{abbrev}
\ee
where the prime means derivative with respect to $\vf_0$. 
The corresponding Hamiltonian reads
\bea
\ch &=& N\, \left(\ch_{0} + \sum_{k\ne 0} \ch_k \right)\nn \\
\ch_{0} &=& \frac{1}{K}\, e^{\f}\,\left(\frac{1}{2}\,\pi_{\a}^2
         -  \frac{3}{2}\,\pi_{\f}^2 \right)
         +  \frac{K}{6}\, e^{-\f}\, V_0  \label{hamiltonian}    \\
\ch_{k} &=& -\left[\,\frac{3}{K}\, e^{\f} \left(
|\pi_{k} |^{2} + \pi_{\phi}\,\left(
\vf_{k}\pi_{k} + \vf_{k}^{\ast}\pi_{k}^{\ast}\right)
+ \pi_{\phi}^{2}\, |\vf_{k} |^{2} \right)
+ \frac{K}{3}\, e^{-\phi}\, U_{k} \,|\vf_{k} |^{2} \right] \nn \ ,
\eea 
with the conjugate momenta defined by
\bea
\pi_{\a} &=& \frac{\pa\cl}{\pa\dot{\a}}  =  \frac{K}{N}\, e^{-\phi}\,
             \dot{\a} \nn \\
\pi_{\f} &=& \frac{\pa\cl}{\pa\dot{\f}}  =   \frac{K}{3N}\, e^{-\phi}\,
             \left[\, -\dot{\f} + \sum_{k\not= 0}
             \left(\vf_{k}\dot{\vf}_{k}^{\ast} 
                 + \vf_{k}^{\ast}\dot{\vf}_{k}\right) \right] \label{momenta}\\
\pi_{k} &=& \frac{\pa\cl_{k}}{\pa\dot{\vf}_{k}} = -\frac{K}{3N}\, e^{-\f}\,
             \left(\,\dot{\vf}_{k}^{\ast} - \dot{\f}\,\vf_{k}^{\ast}\right)
             \nn \ .
\eea
The classical background solutions for $V_0 =0$ and the gauge $N=1$ 
(which we adopt from now on) are given by~\cite{pbb}
\be
\a(t) = \a_0 \mp\frac{1}{\sqrt{3}}\,\ln |t| ,\hspace{4mm}
\f = \f_0 - \ln |t| \ ,
\label{sol2}
\ee
with arbitrary integration constants $\a_0$, $\f_0$.
Here and in the following formulae the upper sign refers to the pre--big--bang
branch (($+$) branch) and the lower sign to the post--big--bang branch
(($-$) branch). Since we are interested in the solutions which describe an
expanding universe we should choose the time ranges $t < 0$ for the ($+$)
branch and $t > 0$ for the ($-$) branch. At $t\ra\mp 0$ the solution runs into
the curvature singularity mentioned in the introduction.

Another problem with the above solution is that the dilaton enters the strong
coupling region as soon as the ($-$) branch is reached. This can be avoided
by adding a dilaton potential which stabilizes the dilaton at a sufficiently
small value. Analytic background solutions are, however, difficult to find for
a nontrivial potential making an explicit study of fluctuations very hard.
In practice we will work with the background~\refs{sol2} pretending that
an additional dilaton potential does not change our conclusions dramatically.\\

To find the solutions of the background Wheeler--de Witt equation the
following variables turn out to be useful~:
\be
d_{\pm}=\frac{1}{\sqrt{6}}\,\left(\f\mp\sqrt{3}\,\a\right) \ .
\label{be}
\ee
They are simply exchanged under SFD
\begin{displaymath}
\begin{array}{ccc}
 & \scriptstyle{SFD} & \\
 d_{+} & \longleftrightarrow & d_{-}
\end{array} \ ,
\label{sfd_trafo}
\end{displaymath}
and the classical trajectories~\refs{sol2} are clearly described by
$d_{\pm} =$ const.


\section{Background wave function}

In this section the theory is studied in the minisuperspace $\{\a ,\f\}$ 
assuming a vanishing dilaton potential. In terms of the variables $d_{\pm}$
introduced above the background Hamiltonian takes the simple form
\be
\ch_0 =-\frac{e^{\f}}{K}\,\pi_{d_{+}}\,\pi_{d_{-}} \ .
\label{Hback}
\ee
Then the general solution of the Wheeler--de--Witt equation
$\hat{\ch_0}\,\Psi_0 =0$ can immediately be written in the form
\be
\Psi_0 (d_{+}, d_{-})= \Psi_0^{(+)} (d_{+}) + \Psi_0^{(-)} (d_{-}) \ ,
\label{psi1}
\ee 
with arbitrary functions $\Psi_0^{(\pm )}$.
Having in mind the classical meaning of $d_\pm$ one would like to interpret
the two terms on the RHS of eq.~\refs{psi1} as corresponding to the two
classical branches of the theory. To make this more precise
let us concentrate on semiclassical wave functions
\be
\Psi_0^{(\pm)}(d_\pm ) = e^{i\, S_{\pm}(d_\pm )} \ .
\label{psi2}
\ee
Such a wave function is peaked about a set of classical
trajectories~\cite{hall_corr} specified by
\be
\pi_{\f}=\frac{\pa S_{\pm}}{\pa\f}, \hspace{4mm}
\pi_{\a}=\frac{\pa S_{\pm}}{\pa\a} \ ,
\label{momenta2}
\ee
where $\pi_{\f},\pi_{\a}$ denote the classical expressions for the momenta
in eq.~\refs{momenta}. Indeed, it can be shown that the eqs.~\refs{momenta2}
describe a subset of the classical trajectories~\refs{sol2} 
for~\footnote{The constant $K$ has been absorbed by a shift of the dilaton.}
\be
S_{\pm} = \pm\frac{e^{-\f_0}}{3}\,\left(\f\,\mp\sqrt{3}\,
          \a\right)
        = \pm\sqrt{\frac{2}{3}}\, e^{-\f_0} \, d_{\pm} \ .
\label{classaction}
\ee
This subset is specified by a fixed value of $\f_0$ and can be parameterized
by $\a_0$. Notice that these wave functions correspond to the asymptotic
expressions which were found in ref.~\cite{ven}.\\

We would like to briefly comment on the range of validity of the above
solutions. We are dealing with the one--loop string effective action.
Thus we have neglected higher curvature terms like e.~g.~an $R^2 $--term in
the action. A typical contribution to the Hamiltonian originating from
such a term is
\be
\d\ch \sim e^{3\f} \,\pi_{\a}^4 \ .
\label{deltaha}
\ee
If we require the action of this operator on the solutions~\refs{psi2} to
be negligible as compared to the one--loop terms in the Hamiltonian,
i.~e.~$\d\hat{\ch}\;\Psi_0^{(\pm)} \ll \hat{\ch}_{0,kin}\,\Psi_0^{(\pm)}$
we find the condition
\be
 \f \ler \f_0\ ,
\label{validity}
\ee
or, equivalently, $H\ll 1$ via the classical solutions.
The condition~\refs{validity} restricts the region in minisuperspace
$\{\a ,\f \}$ accessible to an analysis based on the one--loop effective
action (cf.~fig.~1). Any discussion which relies on the wave
function~\refs{psi2} should therefore be confined to that region.
               

\section{Wave function of the dilaton modes}

As a next step we will determine the wave function of the inhomogeneous
dilaton modes for both branches. This will be done on the background specified
by the semiclassical wave functions~\refs{psi2} with $S_\pm$ given by
eq.~\refs{classaction}. Since we consider the model up to linear terms
in the dilaton perturbations only we can start with the following product
Ansatz for the wave function~:
\be
\Psi^{(\pm)} (b_{\pm},\vfb ) = \Psi_{0}^{(\pm)} (b_{\pm})\;
       \prod_{k\ne 0}\,\Psi_{k}^{(\pm)} (b_{\pm}, \vf_{k}) \ .
\label{prod}
\ee
Here the modes $\{\vf_k\}$ have been collectively denoted
by $\vfb$. It can be shown~\cite{hh} that on the given background
$\Psi_0^{(\pm )}$ each wave function $\Psi_k$ fulfills a Schr\"odinger
equation  
\be
\ch_{k} \, \Psi_{k} = i\,\frac{\partial}{\partial t}\, \Psi_{k} \ ,
\label{schroedinger}
\ee
with the following substitutions made in the interaction
Hamiltonian~\refs{hamiltonian}~:
\be
\pi_{\f}\lra\frac{\pa S_{\pm}}{\pa\f}, \hspace{4mm}
\pi_{\a}\lra\frac{\pa S_{\pm}}{\pa\a} \ .
\label{momenta3}
\ee
The time parameter in eq.~\refs{schroedinger} cannot be viewed as an
independent coordinate but is given in terms of the minisuperspace variables.
A solution can be found via the Ansatz
\be
\Psi_{k} = A_{k}(t)\,e^{-B_{k}(t)\, |\vf_{k} |^{2}} \ ,
\label{ansatz}
\ee
which after some manipulations leads to the following set of differential
equations~\footnote{To simplify the notation we will drop the mode number
$k$ from now on whenever possible.}~:
\be
 \ba{lll}
  \dot{A} &=& 3\, i\, e^{\phi}\,A\,B \nn \\
  \dot{B} &=& 3\, i\, e^{\phi}\,B^{2} -6\, e^{\phi}\,\pi_{\phi}\,B
              -i\,\left( 3\, e^{\phi}\,\pi_{\phi}^{2} + 
               \frac{1}{3}\, e^{-\phi}\, U\right) \ .
 \ea
\label{dglsyst}
\ee
Let us first solve the equation for $B$ which has the structure of a general
Riccati equation. Using the standard substitution
\be
B = \frac{ie^{-\f}}{3}\frac{\dot{D}}{\, D} \ ,
\label{trafo}
\ee
it can be converted into a linear second order differential equation for
$D$~:
\be
\ddot{D} +  \dot{\f}\,\dot{D} + \left(\dot{\f}^2 + U\right)\, D = 0 \ .
\label{secondorder}
\ee
So far we have kept the general form~\refs{abbrev} of $U$. We now
specialize to the case $V_0=0$ or $U = k^2 \, e^{-2\a }$ which corresponds
to the chosen semiclassical background. Then the solution of
eq.~\refs{secondorder} is given by
\be
D^{(\pm )} = |t|\, \cc_0 \left(\frac{e^{-\a_0}\, k}{1\pm\frac{1}{\sqrt{3}}}\;
    |t|^{1\pm\frac{1}{\sqrt{3}}} \right) \ ,
\label{solution1}
\ee
where $\cc_\n $ denotes any linear combination of the Hankel--functions
$H_\n^{(1)}$ and $H_\n^{(2)}$. Inserting this expression in eq.~\refs{trafo}
and reexpressing the time by the background fields we finally arrive at
\be
B^{(\pm )} = \mp\,\frac{i H}{3}\, e^{-\f}\;\left[ 1 - 
          \frac{x}{\sqrt{3}}\,\frac{\cc_{1} (\s^{(\pm )}\, x )}
          {\cc_{0} (\s^{(\pm )}\, x)} \right] \ ,
\label{solutionB}
\ee
with
\be
 x=\frac{k}{aH}\; ,\quad\quad\s^{(\pm )}=\frac{1}{\sqrt{3}\pm 1}\ ,
\ee
and the Hubble parameter $H=\dot{a}/a$.
The prefactor $A$ which appears in the Ansatz~\refs{ansatz} can be partially
fixed by normalizing $\Psi_k$ according to
\be
\int\, \frac{i}{2}d\vf_k \,d\vf^\ast_k \left|\Psi_{k} \right| ^2  =  1 \ .
\label{normpsi}
\ee 
This results in
\be
A = \frac{e^{i\,\b }}{\sqrt{\pi}}\; \left( B + B^{\ast}\right)
    ^{\frac{1}{2}} \ ,
\label{A}
\ee
with an arbitrary phase $\b$, which is determined by the first
equation~\refs{dglsyst}. Since all further considerations do not depend
on this phase we will not compute its explicit form here.

Clearly, for this procedure to be well defined we have assumed that $\Psi_k$
is normalizable at all, i.~e.~$\Re \, (B)>0$. Normalizability of the
inhomogeneous part of the wave function should be required in any case for
a well defined quantum theory of the model in the sense of QFT in curved space.
The explicit expression for $\Re \, (B)$
\be
\Re\, (B) = \frac{e^{-\f}}{3}\,\frac{\Im\, (\,\cc_0 \,\dot{\cc}_0^{\ast}\, )}
           {\left|\cc_0 \right| ^2} \ ,
\label{realB}
\ee
which follows from eq.~\refs{solutionB} shows that this is not automatically
fulfilled but has to be guaranteed by an appropriate choice of the mode
function $\cc_0$. To analyze this in more detail we note that the wave
function does not depend on the overall normalization of $\cc_0$.
Its most general form can therefore be written as
\be
\cc_0 = H_{0}^{(2)} + z\, H_{0}^{(1)} \ ,
\label{vacuum}
\ee 
with an arbitrary complex number $z$. It should be stressed that this quantity
can be chosen independently for each mode $k$. The set of all these
quantities $\{ z_k\}$ fixes the vacuum of QFT in curved space.
From eq.~\refs{vacuum} and the classical trajectories~\refs{sol2} we find
\be
\Im\, \left(\,\cc_0 \,\dot{\cc}_0^{\ast}\,\right) 
     =  \frac{2}{\sqrt{3}\,\p}\,\frac{1}{\s^{(\pm )}\, t} \;
        \left( 1 - |z|^2 \right) \ ,
\label{normCe2}
\ee
which should be strictly positive.
Remember that for an expanding universe the time range was restricted to
$t<0$ for the ($+$) branch and $t>0$ for the ($-$) branch. We
conclude that normalizability of the wave function restricts the vacuum
parameters $z$ to the following regions of the complex plane~:
\be
\ba{lll}
|z| > 1 & {\rm \hspace{4mm} (+) \hspace{1mm} branch} & \\
|z| < 1 & {\rm \hspace{4mm} (-) \hspace{1mm} branch} & . \ea
\label{restriction}
\ee
It should be mentioned that a canonical quantization of the dilaton in the
classical background~\refs{sol2} would lead to similar constraints on its
mode functions once the standard normalization condition with a future oriented
timelike vector is applied~\cite{bd}.


\section{Reduced density matrix and decoherence}

In terms of the wave function the density matrix of a pure quantum system
is given by
\be
\rho_{\ve\tilde{\ve}} \left( b, \tilde{b}, \vfb, 
     \tilde{\vfb} \right) 
   = \Psi^{(\ve )\ast}\,\left( b\, ,\vfb\right)\;
     \Psi^{(\tilde{\ve})}\,\left( \tilde{b}\, ,\tilde{\vfb}\right) \ .
\label{dens}
\ee
Here the quantities $b$, $\tilde{b}$ denote the background variables
and $\ve ,\tilde{\ve} =\pm$ refer to the pre-- and post--big--bang branch.
Not all the degrees of freedom which the pure density matrix depends on
are observed or even observable. The real object of interest is therefore
a reduced density matrix which is obtained from eq.~\refs{dens} by tracing
out certain degrees of freedom. A very natural choice are the inhomogeneous
modes with wavelengths greater than the horizon ($k < aH$) which are clearly
unobservable. Moreover, this choice which we adopt here introduces an
ultraviolet cutoff which serves as a simple regularization of the infinite
product in eq.~\refs{dens}. Let us denote the modes inside the horizon
($k > aH$) by $\cb$. Then the reduced density matrix can be written as
\be
\rho_{\ve\tilde{\ve}}^{({\rm red})} \left( b, \tilde{b}, \cb , 
      \tilde{\cb} \right)  =    
     {\Psi^{\ve}}^{\ast}\left( b\, ,\cb\right)\;
      \Psi^{\tilde{\ve}}\left( \tilde{b}\, ,\tilde{\cb} \right) 
      \;\D^{(\ve\tilde{\ve})}\left( b, \tilde{b} \right) \ .  
\label{reddens}
\ee
The additional factor $\D$ appears as a consequence of the tracing procedure
and can be written as
\be
\D^{(\ve\tilde{\ve})}\left( b, \tilde{b} \right) = \prod_{k\, <\,aH}\,
\D^{(\ve\tilde{\ve})}_k \left( b, \tilde{b} \right) \ ,
\label{Delta}
\ee
with
\be\ba{lll}
\D^{(\ve\tilde{\ve })}_k \left(b, \tilde{b} \right) & = & 
\displaystyle{\int}\frac{i}{2}\, d\vf_k\, d\vf^\ast_k\, 
\Psi^{(\ve)\ast}_k\left( b\, ,\vf_{k} \right) \;
\Psi^{(\tilde{\ve })}_k
\left(\tilde{b}\, ,\vf_{k} \right) \\[0.3cm]
 & = & e^{-i\,\b_k^{(\ve)}(b) 
       + i\,\b_k^{(\tilde{\ve})}(\tilde{b})}\;\;
       \displaystyle{
       \frac{\left(B_k^{(\ve)}\, (b) + B^{(\ve )\ast}_{k}\, (b)\right)
       ^{\frac{1}{2}} \,
       \left(B_k^{(\tilde{\ve})}\, (\tilde{b}) 
       + B^{(\tilde{\ve} )\ast}_{k}\, (\tilde{b})\right)
       ^{\frac{1}{2}}}{B^{(\ve )\ast}_k\, (b) + 
       B^{(\tilde{\ve})\ast}_k\, (\tilde{b})}} \ .
\label{DeltaKa}
\ea\ee
In the second step we have used the Ansatz~\refs{ansatz} for the
wave function and eq.~\refs{A}. Using the shorthand notation
$\D^{(\ve\tilde{\ve})}_k(b,\tilde{b})\rightarrow \D$,~
$B^{(\ve )}_k\, (b) \ra B$ and 
$B^{(\tilde{\ve} )}_k\, (\tilde{b}) \ra \tilde{B}$ we find
\be
|\D |^2 = \left[ 1 + \frac{1}{4}\frac{\left|\, B - \tilde{B}\,\right|^2}
          {\Re \, (B) \,\Re \, (\tilde{B})} \;\right]^{-1} \ .
\label{DeltaKa2}
\ee

A suppression of the off--diagonal entries in
$\r^{({\rm red})}$, i.~e.~decoherence is achieved if $|\D | <1$ for a
sufficiently large number of modes $k$. Conversely, loss of decoherence
occurs under two circumstances~:
\begin{itemize}
\item if the number of modes to be traced out becomes small or
\item if $B=\tilde{B}$ for almost all unobserved modes.
\end{itemize}
The first option is realized if $aH=\exp (\a +\a -\f_0 ) = O(1)$, i.~e.~in
the region of minisuperspace specified by
\be
 \a +\f \ler \f_0\; .
\ee
As can be seen in fig.~1 every ($+$) trajectory emerges from this region and
every ($-$) trajectory will finally end up there. Classical
behaviour is therefore lost for $t\rightarrow\mp\infty$ though the curvature
corresponding to the classical trajectories is small in this limit.

The second possibility for loss of decoherence ($B=\tilde{B}$) is somewhat more
difficult to ana\-lyze. From eqs.~\refs{solutionB} and~\refs{vacuum} we find
that $B$ can be written as a GL$(2,\CC )$ transformation which acts on the vacuum parameter $z$~: 
\be
B = \frac{az - a^{\ast}}{cz + c^{\ast}} \equiv T\, z \ ,
\label{GL2C}
\ee
with
\bea
a &=& \mp\frac{i}{3}\, e^{-\f}\, H\,
      \left(\sqrt{3}\, H_0^{(1)}\, (\s^{(\pm)}\, x) - 
      x\, H_1^{(1)}\, (\s^{(\pm)}\, x)\right) \nn \\
c &=& H_0^{(1)}\, (\s^{(\pm)}\, x)  \ . 
\label{entries}
\eea
Indeed, we have
\be
\det T = 2\, \Re\, (ac^{\ast} ) = \pm\frac{4\, e^{-\f}\, H}
         {3\p\s^{(\pm)}} \not= 0 \ ,
\label{determinant}
\ee
which shows that $T$ can be inverted. Let us first discuss a superposition 
of two semiclassical contributions 
$\Psi_0^{(\e )}+\tilde{\Psi}_0^{(\tilde{\e} )}$ which can be of the same
or different branch type. As we have seen, both wave functions describe a
set of classical trajectories parameterized by $\a_0 $. We would like to
discuss the coherence condition for two of those trajectories, one out of
each set. The vacua which can be chosen independently are denoted by 
$\{ z_k\}$ and $\{ \tilde{z}_k\}$. It is now useful to write the coherence
condition $B=\tilde{B}$ in the form
\be
 z=S_{x,\tilde{x}}\, \tilde{z}\ ,\label{coh_cond}
\ee
with $S_{x,\tilde{x}}=T^{-1}\tilde{T}$. We have indicated the dependence of
$S$ on the background and the mode number through $x=k/aH$. Note that for
the relevant modes $k$ we always have $0<x,\tilde{x}<1$. Given the
restrictions~\refs{restriction} in the complex plane it is not obvious that
for a fixed vacuum $\tilde{z}$ eq.~\refs{coh_cond} leads to a meaningful
$z$. It can, however, be shown that $S$ is an automorphism of the
unit disc $\{ |z|<1\}$ if we consider backgrounds corresponding to the
same branch, i.~e.~$(\ve ,\tilde{\ve})=(\pm ,\pm )$. In the opposite case
$(\ve ,\tilde{\ve})=(\pm ,\mp )$ $S$ maps the unit disc into its complement
in $\CC^\ast$ and vice versa.
Consequently, for any given vacuum $\tilde{z}$ in a semiclassical background
$\tilde{\Psi}_0^{(\tilde{\e})}$ we can find 
-- via eq.~\refs{coh_cond} -- a well defined
vacuum $z$ in the background $\Psi_0^{(\e )}$ such that coherence 
between the trajectories specified by $x$, $\tilde{x}$ occurs.

The situation is somewhat more restrictive if coherence of two trajectories
described by a single semiclassical wave function 
$\Psi_0^{(\e )}$ with a vacuum $\{ z_k\}$ is discussed. 
Then the coherence condition reads $z=S_{x,\tilde{x}}\, z$. 
Since in this case $S_{x,\tilde{x}}$ is an automorphism
of the unit disc and possesses exactly two fix points (one inside and one
outside the unit disc) a solution exists if the
vacuum is chosen to coincide with one of these fix points.


\begin{figure}[h]
\renewcommand{\baselinestretch}{1.1} \small \normalsize 
     \centerline{\psfig{figure=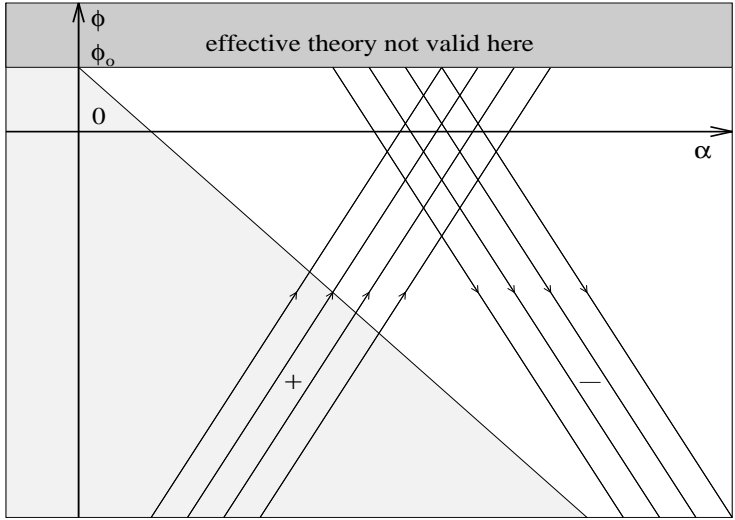,height=70mm,width=100mm}}
     \caption{The minisuperspace $\{\a ,\f\}$ is depicted. The lines
              correspond to classical $(\pm )$ trajectories with the 
              arrows indicating
              the arrow of time. The effective theory breaks down in the
              dark shaded region. In the shaded region the number of
              beyond--horizon modes becomes small.}
     \label{fig1}
     \par \renewcommand{\baselinestretch}{1.5}
\end{figure}

\begin{figure}[h]
\renewcommand{\baselinestretch}{1.1} \small \normalsize 
     \centerline{\psfig{figure=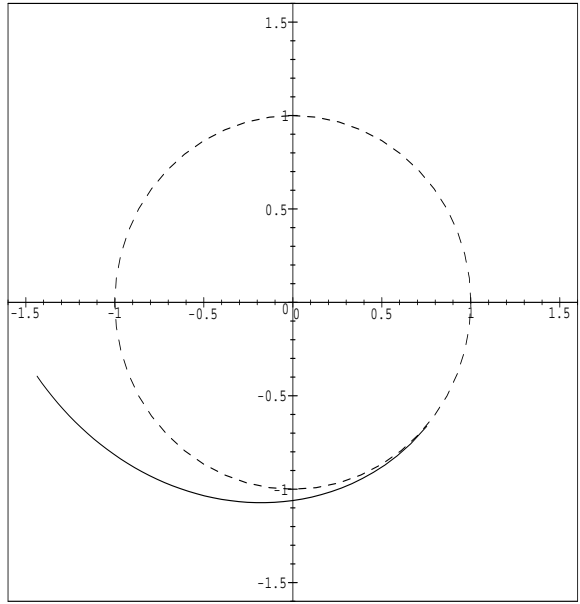,height=70mm,width=110mm}}
     \caption{The complex $z$--plane of vacua splitted into ($+$) 
              and ($-$) regions by the circle of radius 1 is shown. 
              The additional line represents the $(+)$
              image of the $z^-=0$ vacuum for $0<x=k/aH<1$.
              A $z^+$ vacuum on this line can lead to coherence between
              a $(+)$ and a $(-)$ trajectory.}
     \label{fig2}
     \par \renewcommand{\baselinestretch}{1.5}
\end{figure}

                 
\section{Summary and conclusion}
In this paper we have analyzed the quantum cosmology of the simplest
pre--big--bang model with particular emphasis on the aspect of decoherence.

It has been shown that in minisuperspace (spanned by $\{\a =\ln a,\f\}$ with
the scale factor $a$ and the invariant dilaton $\f$) semiclassical wave
functions $\Psi_0^{(\pm )}$ which correspond to the pre--big--bang ($+$)
and post--big--bang ($-$) trajectories can be found. More
precisely, these wave functions single out a one--parameter set of classical
trajectories specified by a fixed value of $\f_0$ and parameterized by $\a_0$
($\a_0$, $\f_0$ are the values of $\a$, $\f$ at $|t|=1$, i.~e.~at the point
where the trajectories enter the high curvature region).\\

On the backgrounds specified by these minisuperspace wave functions we have
determined the wave function of the inhomogeneous dilaton modes and we have
argued that tracing out the modes $k<aH$ beyond the horizon $H^{-1}$ may
cause decoherence of classical trajectories.

Decoherence is lost under two circumstances~: if the number of modes $k$
beyond the horizon is small or for specific choices of vacua (in the sense
of QFT in curved space). We would now like to discuss these conditions and
their implications in more detail.

Let us begin with the first one which results in a purely kinematic
condition and can be discussed in minisuperspace. Schematically, the
situation in minisuperspace has been depicted in fig.~1. We have drawn a
number of ($+$) and ($-$) trajectories specified by $\f\mp\sqrt{3}\a =$ const.
The arrows indicate the direction of time for an expanding universe. For
$t\rightarrow\mp 0$ all trajectories enter the high curvature regime
$\f >\f_0$ (dark shaded region) which is not accessible to our analysis based
on a one--loop effective action. The number of beyond--horizon modes
becomes small for $\f +\a \ler\f_0$, i.~e.~in the shaded region of fig.~1.
All trajectories run into this region for $t\rightarrow\mp\infty$ where
classical behaviour is expected to break down despite a small value of
the curvature. For the ($-$) trajectories this corresponds to the well--known
fact that more and more modes return into the present horizon. In the white
edge of fig.~1 there might exist a well defined notion of classical
trajectories emerging from the quantum cosmological description. Whether or
not this is the case, i.~e.~whether or not decoherence occurs crucially
depends on the chosen vacua. This leads to the second condition
mentioned above.\\

As an example, we would like to discuss a minisuperspace wave function
$\Psi_0 =\Psi_0^{(+)}+\Psi_0^{(-)}$ which is a superposition of two
semiclassical wave functions $\Psi_0^{(\pm )}$ peaked about sets of pre--
and post--big--bang trajectories, respectively. This wave function, as it
stands, cannot describe a classical behaviour since it corresponds to a
superposition of macroscopically different states.

The vacua for both types of trajectories can be parameterized by complex
numbers $\{ z^{(\pm )}_k\}$. We have seen that for the fluctuation wave
function to be normalizable the restrictions $|z^{(+)}|>1$ and
$|z^{(-)}|<1$ should hold. Furthermore, the coherence condition can be written
as $z^{(+)}=S_{x^+,x^-}\, z^{(-)}$ with a GL$(2,\CC )$ map $S_{x^+,x^-}$ which
depends on $x^{\pm}=k/(aH)^{(\pm )}$. The latter quantities specify the ratio
of the horizon length to the wave length for a given mode $k$. For the modes
outside the horizon which are of relevance here we have $0<x^\pm <1$.
It turned out that despite the constraints on $z^\pm$ the decoherence
condition always has a well defined solution. This is illustrated in fig.~2
where the image of the vacuum $z^-_k=0$ (a choice corresponding to the
second Hankel function $H^{(2)}$ which for small wave length reduces to the
standard mode functions in Minkowski space) under
$S_{x,x}$ for $0<x<1$ is displayed~: the whole line of image
points rests outside the unit disc. Clearly, there is no chance for coherence
if the vacuum $\{ z^+_k\}$ is not located on this line. Conversely, if the
$\{ z^+_k\}$ are placed on the line in an appropriate way coherence may occur,
however, for a specific value of the background quantity $aH$ only. This
can serve as a mechanism for branch change which takes place precisely when
the macroscopic expansion reaches the given value of $aH$ and decoherence
between the two branches is lost. Though we have considered the special
case $x^+=x^-$ and a $k$--independent ($-$) vacuum a similar discussion
applies to the general case. Furthermore, if the minisuperspace wave
function is a superposition of two semiclassical contributions corresponding
to the same branch the above conclusions holds analogously.

The coherence condition for two trajectories described by a single
semiclassical wave function is given by the fix point equation
$z=S_{x,\tilde{x}}\, z$ which has a uniquely defined solution $z$ for both
branches. If most of the parameters $\{ z_k\}$ do not coincide with one
of these fix points the trajectories decohere.\\

To summarize, we have shown that decoherence of two trajectories with
arbitrary branch type depends on the choice of vacua but will occur in
the ``generic'' case as long as the region of minisuperspace indicated
by the white edge in fig.~1 is considered. 
This statement holds for trajectories emerging
from the same semiclassical wave function as well as for trajectories
corresponding to different semiclassical backgrounds.
Therefore classical behaviour can be usually assumed in this region.
This also implies that an overlap of the ($+$) and ($-$) wave functions
$\Psi_0^{(\pm )}$ in minisuperspace which might be a source of branch change
is generically destroyed once inhomogeneous modes are included.
Specific vacua, however, lead to coherence e.~g.~ between the ($+$) 
and the ($-$) branch at a certain point in minisuperspace (which can be a
crossing points of the trajectories in fig.~1). This mechanism can generate a
branch change. It should be stressed that such a situation though it can be
arranged in principle does not appear to be very natural unless a reason for
this specific choice of vacua has been found. Unfortunately, we are not aware
of such a reason.\\

In this paper, we have confined ourselves to the simplest case of vanishing
dilaton potential. It would be clearly desirable to generalize the discussion
in that respect and include a simple potential which can stabilize the
dilaton in the perturbative region. It has been shown~\cite{bru_ven} that
in regions of negative potential values close to the ``egg'' a classical
branch change potentially occurs but in fact does not because of dynamical
reasons. It would be interesting to study the decoherence behaviour close
to these specific points in minisuperspace.\\


{\bf Acknowledgment} This work was partially supported by the EC under
contract no.~SC1-CT92-0789 and by the Sonderforschungsbereich 375--95
``Research in Astroparticlephysics'' of DFG.


%
\end{document}